\documentclass[10pt,conference]{IEEEtran}
\usepackage{epsfig}
\usepackage{times}
\usepackage{float}
\usepackage{afterpage}
\usepackage{amsmath}
\usepackage{amstext}
\usepackage{amssymb,bm}
\usepackage{latexsym}
\usepackage{color}
\usepackage{array}
\usepackage{makecell}
\usepackage{diagbox}
\usepackage{amsmath}
\usepackage{amsthm}
\usepackage{graphicx}
\usepackage[center]{caption}
\usepackage{pstricks}
\usepackage{subfigure}
\usepackage{booktabs}
\usepackage{enumerate}
\usepackage{algorithm,algpseudocode}
\usepackage{stackengine}

\def\delequal{\mathrel{\ensurestackMath{\stackon[1pt]{=}{\scriptstyle\Delta}}}}
\usepackage[left = 0.7in, right=0.7in, top=0.75 in, bottom=0.75in]{geometry}

\usepackage[normalem]{ulem}
\usepackage{url}
\newtheorem{theorem}{Theorem}
\newtheorem{lemma}{Lemma}
\newtheorem{definition}{Definition}

\allowdisplaybreaks

\makeatletter
\def\BState{\State\hskip-\ALG@thistlm}
\makeatother

\usepackage{graphicx} 

\title{On Secrecy Capacity of Binary Beampointing Channels with Block Memory and Feedback}
\author{ \IEEEauthorblockN{Siyao Li\IEEEauthorrefmark{1}, 
Mingzhe Chen\IEEEauthorrefmark{2},
     Shuangyang Li\IEEEauthorrefmark{3}, and
     Giuseppe Caire\IEEEauthorrefmark{3}} \\
\IEEEauthorblockA{\IEEEauthorblockA{\IEEEauthorrefmark{1} Electrical Engineering and Computer Science Department, Embry-Riddle Aeronautical University, Daytona Beach, FL, USA}
\IEEEauthorblockA{\IEEEauthorrefmark{2} Electrical and Computer Engineering Department, University of Miami, Coral Gables, FL, USA}
\IEEEauthorrefmark{3}Faculty of Electrical Engineering and Computer Science, Technical University of Berlin, Berlin, Germany}
\IEEEauthorblockA{E-mail: lis14@erau.edu, mingzhe.chen@miami.edu,  shuangyang.li@tu-berlin.de, caire@tu-berlin.de}
}

\begin{document}

\maketitle

\begin{abstract}
    This paper investigates the secrecy capacity of the binary beampointing (BBP) channel with block memory and feedback, a simplified yet insightful model for millimeter-wave (mmWave) systems with beamformed transmissions and backscatter feedback. We consider a system where a legitimate receiver and a passive eavesdropper experience independent and uniformly distributed angular directions over transmission blocks, with the base station receiving noiseless, unit-delayed feedback from both, under the per-symbol input cost constraints. 
    We establish a closed-form upper bound on the secrecy capacity, which is based on the main channel between the base station and the legitimate receiver. Moreover, we propose a joint communication and adaptive sensing (JCAS) scheme and derive its achievable secrecy rate. Simulation results show that the gap between the inner and outer bounds narrows as the number of block length increases. This reveals the efficiency of this JCAS scheme, which strategically leverages feedback to balance the demands of sensing the legitimate user and preventing information leakage to the eavesdropper.  
\end{abstract}

\section{Introduction}
Sixth-generation (6G) wireless systems are envisioned to operate at higher frequency bands, including millimeter-wave (mmWave) frequencies, to support unprecedented data rates and a wide range of new applications \cite{6GTerahertz}. However, communication at mmWave frequencies is highly susceptible to severe pathloss and blockages, necessitating the use of highly directional beams through advanced beamforming techniques  \cite{hong2021role}. The initial phase of establishing reliable communication in such scenarios involves a crucial beam acquisition process, where the transmitter must identify the optimal beam direction towards the receiver \cite{Michelusi_2018}. 
To maximize spectral efficiency and resource utilization in 6G, there is a growing interest in integrating sensing and communication functionalities \cite{ISAC-6G,EnableJCAS}. This paradigm allows communication waveforms to simultaneously perform environmental sensing, such as target detection and localization, alongside data transmission \cite{ISAC-IT}. In scenarios where the receiver’s location is unknown, jointly incorporating initial beam probing (sensing) with the communication phase presents a promising strategy to achieve efficient beam alignment while maintaining data transmission capabilities \cite{Li2022Asilomar,Li2023ISIT,BBP:TIT2023}. 

This work investigates the fundamental limits of secure communication within a binary beampointing (BBP) channel model \cite{BBP:TIT2023}, which serves as a simplified yet insightful representation of the initial beam acquisition phase in directional communication systems. We consider a system comprising a legitimate user and a passive eavesdropper. To reflect the quasi-static nature of user and eavesdropper positions over short periods and their potential mobility, we model the channel states, i.e., Angle-of-Departure (AoD) directions, as remaining constant over blocks of symbols, i.e., in-block memory (iBM), and changing independently and identically distributed (i.i.d.) across different blocks. Prior work has shown that feedback signals play a critical role in stability \cite{Huang2012TIT,Li2019ITW}, capacity \cite{Li2019ICC,Tatiknonda2009,Li2021ISIT} and state estimation \cite{caire2007multiuser,Li2023ITW}. We incorporate the practical aspect of feedback, assuming that the base station (BS) receives backscatter signals from both the legitimate user and potentially the eavesdropper, which can aid in refining the estimates of their positions. 

The problem of secure communication in the presence of an adversarial eavesdropper has been extensively studied within the framework of information theory, initiated by Wyner's seminal work on the wiretap channel \cite{wiretap}. Subsequent research has explored the secrecy capacity for various discrete memoryless channel models, including broadcast channels with confidential messages \cite{Csiszar-Korner}, fading channels \cite{secrecy-fading-2008,wang2007secrecy}, multi-antenna systems such as multiple-input single-output (MISO) \cite{shafiee2007achievable,secureMISOME2010} and multiple-input multiple-output (MIMO) wiretap channels \cite{yeh2021eavesdropping}, as well as channels with feedback \cite{lai2008wiretap,bashar2011secrecy}. The impact of feedback on secrecy has also been investigated for unifilar finite-state channels \cite{permuter2008capacity} and memoryless state-dependent broadcast channel under joint communication and sening (JCAS) scenario \cite{gunlu2023secure}. However, existing studies often rely on traditional metrics and may not fully address the unique aspects of emerging paradigms or the interplay between different system components in terms of secrecy.  Challenges remain in analyzing complex systems, such as those beamforming systems with iBM and feedback. 

\paragraph*{Contributions} In this work, we adopt and extend the BBP channel model introduced in \cite{BBP:TIT2023}, which captures the key characteristics of block-based beam alignment and feedback-aided communication in a tractable framework. We present the information-theoretic secrecy analysis of the BBP channel with feedback and iBM. The key novelty of our work lies in (i) formalizing a secrecy capacity framework for this new class of JCAS-inspired channels, (ii) deriving an upper bound on the secrecy capacity based on feedback-assisted channel capacity, and (iii) establishing a lower bound on the achievable secrecy rate via a difference-of-mutual-information formulation between the legitimate and eavesdropper links. Additionally, we propose a practical JCAS transmission strategy that leverages feedback to approach the secrecy bounds under realistic per-symbol input cost constraints.

\paragraph*{Organization} The remainder of this paper is organized as follows. Section \ref{sec:system} introduces the system model, necessary definitions and capacity results. Section \ref{sec:main} presents the main results. Section \ref{sec:simulation} evaluates the performance with numerical examples. Section \ref{sec:conclusion} concludes this work. 

\paragraph*{Notations}
For an integer $n$, we let $[n] = \{ 1, \cdots, n\}$ and $[n_1: n_2] = \{ n_1, \cdots, n_2\}$ for some integers $n_1 < n_2$.  $\underline{X}$ denotes a vector and $\underline{X}^n = [\underline{X}_1, \cdots, \underline{X}_n ]$ denotes a sequence of vectors. 
Let $H( \cdot )$ denote the binary entropy function. We take $\log$ with base 2 throughout the paper.  
\begin{figure}[t]
\centerline{ \includegraphics[width=0.65\linewidth]{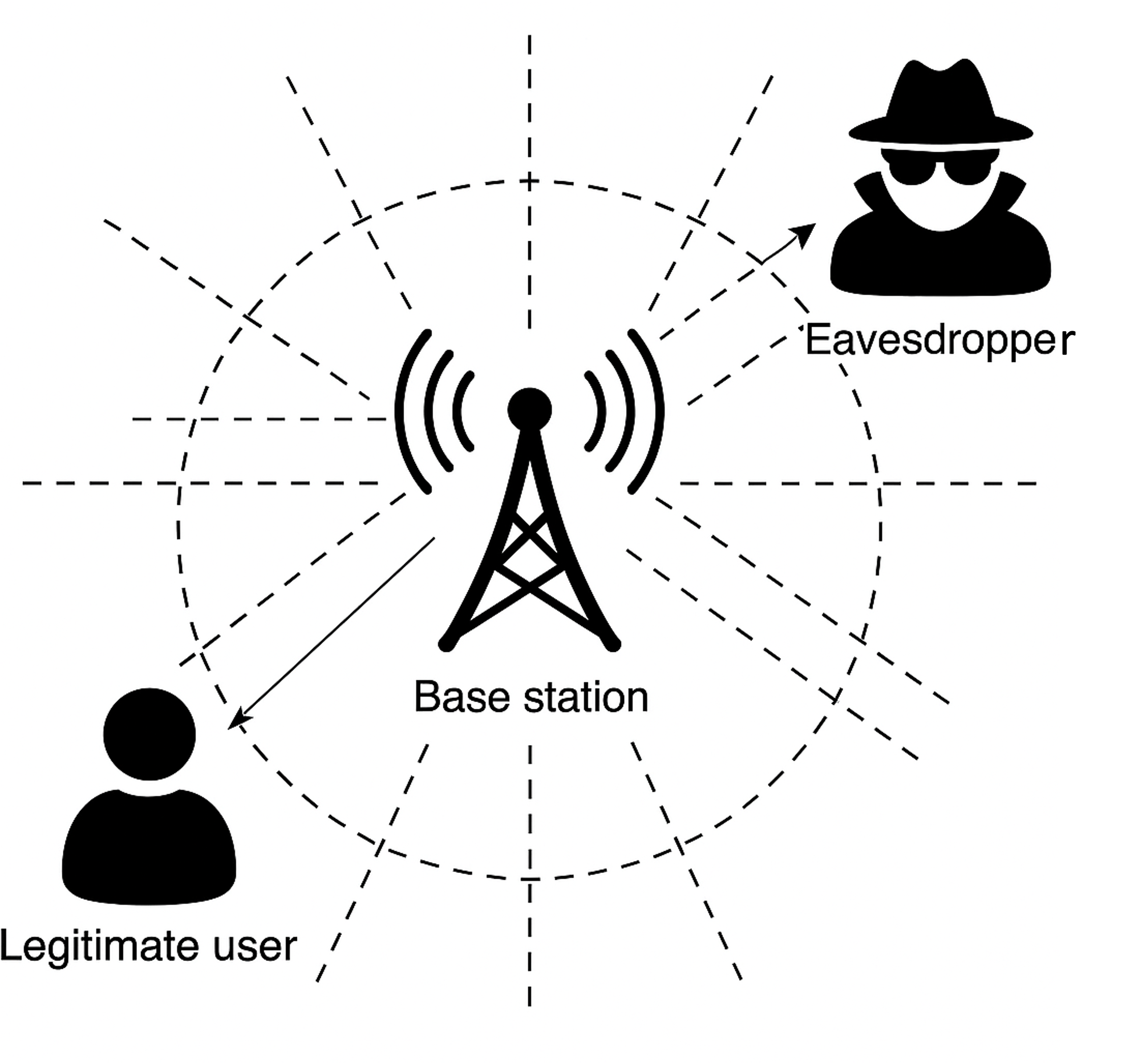} \hspace{.0cm}  }
\caption{System model.}
     \label{fig:model}
	\vspace{-.5cm}
\end{figure}

\section{System Model}
\label{sec:system}
In this work, we focus on the BBP channel model introduced in \cite{BBP:TIT2023} with a legitimate user and an adversarial eavesdropper under  per-symbol input cost constraint $B$ at the transmitter with iBM $L$ and unit-delayed feedback from both receivers (see Fig. \ref{fig:model}). 
The transmitter sends a signal $\underline{X}_{i,j}$ to a receiver based on a message $W \in \mathcal W = \{1, \cdots, M\}$ and the feedback from both the receiver and the adversarial eavesdropper.
At channel use $j$ in $i$-th block, the legitimate user receives
\begin{align}
Y_{i,j}^{(l)} = {{\underline S}_{i}^{(l)}}^T {\underline X}_{i,j} 
\end{align} 
and the eavesdropper receives
\begin{align}
Y_{i,j}^{(e)} = {{\underline S}_{i}^{(e)}}^T {\underline X}_{i,j}, \label{eq:channel-model}
\end{align} 
where ${\underline S}_i^{(l)}, {\underline S}_i^{(e)} \in \mathcal S\delequal \{0, 1\}^K$ are one-hot vectors with the one element being one and zero otherwise, 
$\underline X_{i,j} \in \mathcal X \delequal \{0, 1\}^K$ is the $j$-th input symbol $j \in[L]$ in block $i\in[\ell]$,
and $Y_{i,j}^{(l)}, Y_{i,j}^{(e)} \in \mathcal Y \delequal \{0,1\}$.  
The indices  of the ``one'' element in ${\underline S}_i^{(l)}$, and ${\underline S}_i^{(e)}$ 
indicate the AoD directions from the BS to the legitimate user and adversarial eavesdropper, respectively, and remain constant over the $i$-th block. 
In addition, they are i.i.d. and uniformly distributed over $[K]$ \footnote{The eavesdropper and the legitimate user may locate in the same direction.}. 
The channel state ${\underline S}_i^{(l)}$ is known at the legitimate receiver, and ${\underline S}_i^{(e)}$ is known at the eavesdropper. 


The transmitter receives one-unit-delayed noiseless feedback from both the receiver and the eavesdropper, i.e., $Y_{i,j}^{(l)}$ and $Y_{i,j}^{(e)}$  at the end of channel use $j$ of block $i$ \footnote{Even though the legitimate receiver and the eavesdropper may be located along the same angular direction (i.e., associated with the same beam index), the BS is assumed to be able to distinguish between their feedback signals. This is possible because the legitimate receiver and eavesdropper are not co-located at exactly the same distance from the BS, which leads to distinguishable time-of-arrival delays in the feedback signals. Additionally, practical systems can incorporate small signaling overheads, such as unique pilot sequences or source identification codes, that allow the BS to reliably attribute each feedback signal to the correct node. }. At the end of each block, the transmitter estimates the state $\hat{\underline S}_i^{(l)}$, $\hat{\underline S}_i^{(e)}$ from the channel input ${\underline X}_{i}^L$ and feedback ${Y_i^{(l)}}^L, {Y_i^{(e)}}^L$, where we denote $\underline X_{i}^{L}  = [ \underline X_{i,1}, \cdots, \underline X_{i, L}]$. 
The joint probability distribution of the considered model is 
\begin{align*}
&P_{W {\underline X}^n {{\underline S}^{(l)}}^{\ell}, {{\underline S}^{(e)}}^{\ell},  {Y^{(l)} }^n,  {Y^{(e)} }^n }( w, {\underline x}^n, {{\underline s}^{(l)}}^{\ell}, {{\underline s}^{(e)}}^{\ell},  {y^{(l)} }^n, {y^{(e)} }^n ) 
\notag
\\&=P(w)    \prod_{i=1}^{\ell} \left( P( {\underline s}^{(l)}_i ) P( {\underline s}^{(e)}_i )  \right.
\\&  \qquad \left. \times \prod_{j=1}^L \left(  P(  y_{i,j}^{(l)}|{\underline x_{i,j} },\underline s_i^{(l)}  ) P(  Y_{i,j}^{(e)}|{\underline x_{i,j} },\underline s_i^{(e)} ) \right. \right.
\\& \qquad \left. \left. \times 
P({\underline x}_{i,j} | w, { y_{i}^{(l)}}^{j-1}, { y_{i}^{(a)}}^{j-1} ) \right) \right), 
\end{align*} 
where 
 $n = \ell L$ denotes the total transmission time in channel uses,  $\ell$ is the number of blocks and $L$ is the length of a block in channel uses. 


\subsection{Definitions}
The BS wishes to send the uniformly distributed message $W$ ranging over the set $\mathcal W = \{1, \cdots, M \}$ to the receiver using an $(M,n)$ code consisting of 
\begin{enumerate}
    \item a sequence of causal stochastic encoders $f_{i,j}, i \in [\ell], j \in[L]$ at the source that maps the message $w$ and the received feedback signal $\underline Y^{iL + j -1} = \{ (Y_1^{(l)}, Y_1^{(e)}), \cdots, (Y_{iL+j-1}^{(l)}, Y_{iL+j-1}^{(e)}) \}$ to $\underline X_i$ sent at time $i$, i.e., $\underline X_{i,j} = f_{i,j}(W, \underline Y^{iL+j-1})$  subject to the following per-symbol  cost constraint
\begin{align}
 b( \underline X_{i,j})  \leq B, \label{eq:input-cost}
\end{align}
where $b: \mathcal X \to \mathbb{R}_+$ is an input cost function;
\item a decoding function at the receiver $g: {\mathcal{S}}^{\ell} \times {\mathcal{Y}^{(l)}}^{n} \to \mathcal{W}$, such that $\hat{W} = g( {Y^{(l)}}^n, {{\underline S}^{(l)}}^{\ell})$. The rate of such code is $R$ and
 the error probability is:
$P_e^{(n)} \delequal \frac{1}{M } \sum_{w \in \mathcal W} \mathbb{P}( g({Y^{(l)}}^n, {{ \underline S}^{(l)}}^{\ell}) \neq w | w ).$  
\end{enumerate}

The knowledge that the eavesdropper gets of $W$ from its received signal ${Y^{(e)}}^n$ and CSI ${\underline S^{(e)}}^\ell$ is modeled by 
\begin{align}
    I(W; {Y^{(e)}}^n, {\underline S^{(e)}}^\ell) =     H(W) - H(W| {Y^{(e)}}^n, {\underline S^{(e)}}^\ell)     \label{eq:leak-info}
\end{align}
since the mutual information measures the amount of information that the eavesdropper contains about $W$. The notion of perfect secrecy
captures the idea that whatever are the resources available to
the eavesdropper, they will not allow him to get a single bit of
information. Perfect secrecy thus requires \eqref{eq:leak-info} $=0$, i.e.,  
    $H(W) = H(W| {Y^{(e)}}^n, {\underline S^{(e)}}^\ell)$  
to hold asymptotically as $n$ grows. 
Therefore, the equivocation rate at the eavesdropper is defined as 
\begin{align}
    R_e = \frac{1}{n} H(W |{Y^{(e)}}^n, {\underline S^{(e)}}^\ell)
    \label{eq:Re}
\end{align}
with $0 \leq R_e \leq \frac{1}{n} H(W)$. Clearly, if $R_e$ is equal to the information rate $\frac{1}{n} H(W)$, then \eqref{eq:leak-info} $=0$, which yields perfect secrecy. 
Associated with secrecy is a perfect secrecy rate $R_s$, which is the amount of information that can be sent not only reliably but also confidentially, with the help of a $(2^{nR_s}, n)$ code.
\begin{definition}
\label{def:Re}
\normalfont
A secrecy rate $R_s$ is said to be achievable over the jamming channel with feedback if for 
any $\epsilon >0$, there exists a positive number $n_0$ and a sequence of codes $(M,n)$ such that for all $n \geq n_0$, we have
\begin{align}
    R_s = \frac{1}{n} \log M,
    \ 
    P_e^{(n)} \leq \epsilon,
    \\
    b( \underline X_{i,j})  \leq B, \forall i\in[\ell], j\in [L],
\\
   R_e =  \frac{1}{n} H(W |{Y^{(e)}}^n, {\underline S^{(e)}}^\ell ) \geq R_s -\epsilon.
\end{align}
\end{definition}

\begin{definition}
\label{def:Cs}
\normalfont
    The secrecy capacity with feedback $C_s$ is the maximum rate at which messages can be sent to the destination with perfect secrecy, i.e., 
    \begin{align}
        C_s = \sup_{f_{i,j}, g, i\in[\ell], j\in[L]} \{ R_s: R_s \text{ is achievable} \}.
   \label{eq:Cs}
    \end{align}
\end{definition}
 
We consider any sequence of $(2^{nR_s}, n)$ codes with perfect secrecy rate $R_s$ and equivocation rate $R_e$, such that $R_e > R_s - \epsilon$, with peak input cost constraint less than or equal to $B$ and error probability $P_e^{(n)} \to 0$ as $n \to \infty$. 

\subsection{Capacity Results}

\begin{lemma}
\label{lemma:Ce}
The equivocation rate $R_e$ at the eavesdropper in \eqref{eq:Re} is upper bounded by 
\begin{align}
    C_e = \frac{1}{L} \max_{P_{{ \underline X}^L || \underline Y^{L-1}} \in \mathcal P_{B}  }  \sum_{j=1}^L H(Y_{j}^{(l)} | {Y^{(l)}}^{j-1}, {\underline S^{(l)}} )  \label{eq:Ce}
\end{align}
 where 
  $P_{ {\underline X}^L \| \underline Y^{L-1} } \delequal \prod_{j=1}^L P_{ {\underline X}_j | {\underline X}^{j-1}, \underline Y^{j-1}}$, 
 $\underline Y^j = \{ (Y_1^{(l)}, Y_1^{(e)}), \cdots, (Y_j^{(l)}, Y_j^{(e)}) \}$  and 
       $\mathcal P_{B} = \{ P_{ {\underline X}^L \| \underline Y^{L-1} } | b( \underline X_{j})  \leq B, \forall j \in [L] \}.$
\end{lemma}
\begin{IEEEproof} 
By definition,  
\begin{align}
R_e &= H(W |{Y^{(e)}}^n, {\underline S^{(e)}}^\ell ) \notag 
 \\&=  I(W ; {Y^{(l)}}^n, {\underline S^{(l)}}^\ell ) - I(W ; {Y^{(e)}}^n, {\underline S^{(e)}}^\ell ) \notag
 \\& \qquad +  H(W |{Y^{(l)}}^n, {\underline S^{(l)}}^\ell ) \notag
 \\& \leq I(W ; {Y^{(l)}}^n, {\underline S^{(l)}}^\ell ) - I(W ; {Y^{(e)}}^n, {\underline S^{(e)}}^\ell )  +  n \delta_n  \label{eq:decodable}
 \\ &  \leq  I(W ; {Y^{(l)}}^n, {\underline S^{(l)}}^\ell )   +  n \delta_n  \notag 
 \\& \leq  I( \underline X^n ; {Y^{(l)}}^n | {\underline S^{(l)}}^\ell )   +  n \delta_n  \label{eq:W-X}
 \\& = \sum_{i=1}^\ell \sum_{j=1}^L I(\underline X_{i,j}; Y_{i,j}^{(l)} | {Y_i^{(l)}}^{j-1}, {\underline S_i^{(l)}} ) \label{eq:chain-rule}
 \\& =  \sum_{i=1}^\ell \sum_{j=1}^L H(Y_{i,j}^{(l)} | {Y_i^{(l)}}^{j-1}, {\underline S_i^{(l)}} )
\end{align}
where \eqref{eq:decodable} follows from the Fano's inequality and $\delta_n \to 0$ as $n\to \infty$, 
\eqref{eq:W-X} follows from data processing inequality, \eqref{eq:chain-rule}
follows from chain rule and the channel is i.i.d. across blocks where ${Y_i^{(l)}}^{j-1} = \{ Y_{i,1}^{(l)}, \cdots, Y_{i,j-1}^{(l)}\}$.
\end{IEEEproof}
Therefore, the secrecy capacity is upper bounded by 
\begin{align}
    C_s & \leq C_e.
    \label{eq:RelessCe}
\end{align}

Without the adversarial eavesdropper, the channel capacity 
has been presented as follows
\begin{lemma}[\cite{BBP:TIT2023}]
\label{lemma:BBP-peak-capacity}
The channel capacity of the BBP channel with iBM and feedback under the per-symbol cost constraint is 
\begin{align}
  C &= \frac{1}{L} \max_{P_{{ \underline X}^L || {Y^{(l)}}^{L-1}} \in \mathcal P_{B^{\prime} }  }  \sum_{j=1}^L H( Y^{(l)}_j | {Y^{(l)}}^{j-1}, {\underline S}  ) \label{eq:BBP-capacity}
\end{align}
 where 
  $P_{ {\underline X}^L \| {Y^{(l)}}^{L-1} } \delequal \prod_{i=1}^L P_{ {\underline X}_i | {\underline X}^{i-1}, {Y^{(l)}}^{i-1}}$, 
   and 
      $ \mathcal P_{B^{\prime}} = \{ P_{ {\underline X}^L \| {Y^{(l)}}^{L-1} } | b( \underline X_{j})  \leq B, \forall j \in [L] \}.$
\end{lemma}
It is a fundamental principle of information theory that the secrecy capacity of a channel is less than or equal to the capacity of the main channel (the channel to the legitimate user) without considering secrecy constraints or the eavesdropper. Thus,  
\begin{align}
    C_s & \leq C.
    \label{eq:RelessC}
\end{align}
The difference between $C_e$ in Lemma \ref{lemma:Ce} and $C$ in Lemma \ref{lemma:BBP-peak-capacity} lies in the set of admissible input distributions over which the maximization is performed: $\mathcal P_B$ for $C_e$ and $\mathcal P_{B^{\prime}}$ for $C$. The set $\mathcal P_B$ allows the input distribution to depend on joint feedback (legitimate and eavesdropper outputs), while $\mathcal P_{B^{\prime}}$ allows dependence only on legitimate feedback. Since dependence on joint feedback includes dependence on legitimate feedback, the set $\mathcal P_{B^{\prime}}$ is a subset of $\mathcal P_{B}$.

\section{Main Results}
\label{sec:main}
In this section, we present a closed-form expression of the outer bound of the secrecy capacity and provide a lower bound to the achievable secrecy rate with a JCAS transmission strategy illustrated in Algorithm \ref{alg:JSAC-peak}. 


\subsection{Outer Bound}

When in-block memory length $L=1$, the channel is equivalent to a memoryless channel. By \eqref{eq:decodable}, 
\begin{align}
    R_e & \leq \frac{1}{n} \left( I(W ; {Y^{(l)}}^n| {\underline S^{(l)}}^\ell ) - I(W ; {Y^{(e)}}^n| {\underline S^{(e)}}^\ell ) \right)  +   \delta_n \notag 
    \\& = \delta_n  \to 0 \label{eq:L=1-Re0}
\end{align}
where \eqref{eq:L=1-Re0} holds since the channel is memoryless and $\underline S^{(e)}$ and $\underline S^{(l)}$ are independent and i.i.d. across each time slot.
Therefore, no secret message will be transmitted to the legitimate receiver.

For in-block memory $L \geq 2$, we have the secrecy capacity upper bound as follows.
\begin{theorem}
\label{thm:upper}
The secrecy capacity of the BBP channel with feedback and iBM under the per-symbol input cost constraint $B$ is upper bounded by 
    \begin{align}   
 \frac{1}{L} 
\sum_{j=1}^L  \left( (1 -\frac{ \sum_{k=1}^{j-1} c_{k} }{K}  ) H( \frac{c_{j} }{K - \sum_{k=1}^{j -1} c_k }) 
 +  \frac{\sum_{k=1}^{j-1} c_{k} }{K} \right), \label{eq:peak-each-rate}
  \end{align} where $\sum_{k=1}^{0} c_{k} = 0$, $c_1 = \min( \frac{K}{2}, B)$, and
   \begin{align}
    c_j = \min( \frac{K- \sum_{k=1}^{j-1} c_k }{2}, B ), 1 <j \leq L. \label{eq:b0s-peak}
  \end{align} 
\end{theorem}
\begin{IEEEproof}
    The result follows directly from \eqref{eq:RelessC} and Theorem 2 in \cite{BBP:TIT2023}. 
    \end{IEEEproof}

 

\subsection{Inner Bound}

\begin{theorem}
\label{thm:lower}
The secrecy capacity of the BBP channel with feedback and iBM under the per-symbol input cost constraint $B$ is lower bounded by 
    \begin{align}   
 \frac{1}{L} &
\sum_{j=1}^L  \left( (1 -\frac{ \sum_{k=1}^{j-1} c_{k} }{K}  ) H( \frac{c_{j} }{K - \sum_{k=1}^{j -1} c_k }) 
 +  \frac{\sum_{k=1}^{j-1} c_{k} }{K} \right. \notag 
 \\& \left. -  \frac{K - \sum_{k=1}^{j-1} c_k}{K} H(\frac{c_j}{K} ) \right. \notag 
 \\& \left. - \frac{c_{j-1}  (K - \sum_{k=1}^{j-2} c_k) }{K^2} H( \frac{1}{2 } \frac{c_{j-1}}{K - \sum_{k=1}^{j-2} c_k}  ) \right. \notag 
 \\& \left. - \sum_{k=1}^{j-3} \frac{1}{K}  \frac{c_{k+1}^2}{K^2} (\frac{1}{2})^{2(j-k-2)-1} \right), \label{eq:peak-each-rate}
  \end{align} where  $c_j$ is updated as shown in \eqref{eq:b0s-peak}. 
\end{theorem}

\begin{IEEEproof}
Using random coding and binning, one can show that for any rate 
\begin{align}
  R< \frac{1}{L}\left( I( \underline X^L ; {Y^{(l)}}^L | {\underline S^{(l)}} ) - I( \underline X^L ; {Y^{(e)}}^L | {\underline S^{(e)}} ) \right),  
  \label{eq:inner}
\end{align}
 there exists a coding scheme that achieves reliable communication and perfect secrecy asymptotically. This is essentially the Csiszár–Körner secrecy rate formula \cite{Csiszar-Korner} adapted to the BBP channel with iBM. 
 The rate $\frac{1}{L}( I( \underline X^L ; {Y^{(l)}}^L | {\underline S^{(l)}} ) - I( \underline X^L ; {Y^{(e)}}^L | {\underline S^{(e)}} ) )$  represents the difference between the information reliably delivered to the legitimate user and the information gained by the eavesdropper, conditioned on their respective channel states.   
 
 In \cite{BBP:TIT2023}, we have shown a JCAS algorithm that achieves the capacity in Theorem \ref{thm:upper}. To make things clear, we incorporate feedback from the eavesdropper and refine the achievable scheme in Algorithm \ref{alg:JSAC-peak}. 
   Let ${\mathcal B}_{\underline y^i}$ denote the set of beam indices containing the transmission direction at channel use $i$ when channel output $\underline y^i = \{ (y_1^{(l)}, y_1^{(e)}), \cdots, (y_i^{(l)}, y_i^{(e)}) \}$. Let $\mathcal B_i^e$ denote the set of beam indices to be explored at channel use $i$.   We initialize ${\mathcal B}_{\underline y^0} = [K]$, ${\mathcal B}_{0}^e = \emptyset$, and a sequence of $\{ c_1, \cdots, c_L\}$ iteratively solved by \eqref{eq:b0s-peak}. At the beginning of channel use $i$, we update ${\mathcal B}_{\underline y^i}$ and choose some number of beam indices randomly and uniformly from ${\mathcal B}_{\underline y^{i}}$ based on the casual feedback $\underline Y_{i-1} = (y_{i-1}^{(l)}, y_{i-1}^{(e)})$. Specifically,  we use $k, k\in[L]$ to record the number of channel uses until the transmitter selected the ``right" directions (i.e., $Y_{k}^{(l)}=1$ and ${Y^{(l)}}^{k-1}= 0^{k-1}$). Before that, the transmitter randomly and uniformly chooses $c_i, i\leq k$ beam indices from ${\mathcal B}_{\underline y^{i-1}}$, which is updated by excluding previously selected beams iteratively. 
   After that,  the transmitter randomly and uniformly chooses $\max( \frac{c_k}{2^{i-k}}, 1), i>k$ beam indices from ${\mathcal B}_{\underline y^{i-1}}$. These selected beam indices are stored in set $\mathcal B_i^e$, and the set ${\mathcal B}_{\underline y^{i-1}}$ is updated based on feedback to either exclude the previously selected beams when $Y_i^{(l)} = 0$ or proceed with half of the previously selected beams when $Y_i^{(l)} = 1$. 
 
\begin{algorithm}[t] 
\caption{JCAS Scheme} \label{alg:JSAC-peak}
\begin{algorithmic}[1]
 \State {\bf Initialization:} 

1) Let $\underline Y^0 = (0, 0)$, ${\mathcal B}_{\underline y^0} = [K]$ and  $\mathcal B_0^e = \emptyset$.

2) Given a sequence of $\{ c_1, \cdots, c_L\}$ by \eqref{eq:b0s-peak}.
  \State Randomly and uniformly choose $c_1$ beam indices from ${\mathcal B}_{\underline y^{0}}$, and the selected directions are stored in $\mathcal B_1^e$. 
\State {\bf Recursions:}  
  \For{$ i = 2: L$}  
  	 \If {$ {Y^{(l)}}^{i-1} = 0^{i-1}$}
	  \State ${\mathcal B}_{\underline y^{i}} = {\mathcal B}_{\underline y^{i-1}} \backslash \mathcal B_{i-1}^e$.
  \State Randomly and uniformly choose $c_i$ beam indices from ${\mathcal B}_{\underline y^{i}}$, and the selected directions are stored in $\mathcal B_i^e$.	
	 \Else  
	  \State $k=$ the first time index $i-1$ such that $Y_{i-1}^{(l)} = 1$.
      \If {$ {Y_i^{(l)}} = 0$}
 \State ${\mathcal B}_{\underline y^{i}} =\mathcal B_{k}^e \backslash \mathcal B_{i-1}^e $.
 \State $\mathcal B_{k}^e ={\mathcal B}_{\underline y^{i}}$.
      \Else
       \State ${\mathcal B}_{\underline y^{i}} =\mathcal B_{k}^e$.
	   \EndIf
        \State Randomly and uniformly choose $\max( \frac{c_k}{2^{i-k}}, 1)$ beam indices from ${\mathcal B}_{\underline y^{i}}$, and the selected directions are stored in $\mathcal B_i^e$.

	 	\EndIf
		
%
	%
		
\EndFor
\end{algorithmic}
\end{algorithm}
Notice that the transmission strategy is actually independent of the feedback $Y_i^{(e)}$ from the eavesdropper. 
 As proved in \cite{BBP:TIT2023},  Algorithm \ref{alg:JSAC-peak} achieves the channel capacity of the main channel, that is, 
    $\frac{1}{L} I( \underline X^L ; {Y^{(l)}}^L | {\underline S^{(l)}} ) =  $ \eqref{eq:peak-each-rate}.
Comparing the secrecy capacity inner bound in \eqref{eq:inner} to the outer bound in \eqref{eq:BBP-capacity}, the difference is $\frac{1}{L} I( \underline X^L ; {Y^{(e)}}^L | {\underline S^{(e)}} )$ where
\begin{align}
    &\frac{1}{L} I( \underline X^L ; {Y^{(e)}}^L | {\underline S^{(e)}} ) 
   = \frac{1}{L} H( {Y^{(e)}}^L | {\underline S^{(e)}} ) \notag 
    \\& = \frac{1}{L} \sum_{j=1}^L H( {Y_j^{(e)}} | {Y^{(e)}}^{j-1}, {\underline S^{(e)}} ) \notag 
    \\& = \frac{1}{L} \sum_{j=1}^L \mathbb{E} [H( P_{{Y_j^{(e)}} | {Y^{(e)}}^{j-1}, {\underline S^{(e)}}}(1 | {y^{(e)}}^{j-1}, {\underline s^{(e)}}))].
    \label{eq:H1y}
\end{align}

One can easily verify that when $L=1$ the lower bound to the achievable secrecy rate  
\begin{align*}
    R_1 &= I( \underline X ; {Y^{(l)}} | {\underline S^{(l)}} ) - I( \underline X ; {Y^{(e)}} | {\underline S^{(e)}} )
    \\ &= 
    H(\min( \frac{1}{2},  \frac{B}{M} ) ) - H(\min( \frac{1}{2},  \frac{B}{M} ) ) =0.
\end{align*}
When $L \geq 2$, at channel use $j \in[ 2, L)$, based on the feedback,
\begin{itemize}
\item if $Y_{j-1}^{(l)} = 0,$ it means that the transmission direction is not within the beam indices selected by $\underline X_{j-1}$. We simply exclude these selected beam indices from ${\mathcal B}_{\underline y^{j-1}} $, and update ${\mathcal B}_{\underline y^{j}} = {\mathcal B}_{\underline y^{j-1}} \backslash \mathcal B_{j-1}^e$. Then, we apply the same strategy but with $| {\mathcal B}_{\underline y^{j-1}}| - c_{j-1}$ possible directions to be explored,  which provides
\begin{align}
    P[{Y_j^{(e)}} = 1 | \exists i \in[j-1]: \underline Y_i = (0,1) ] = 0.
    \label{eq:01}
\end{align}
\item if $Y_{j-1}^{(l)} = 1$, the right transmission direction is detected in some known set and the peak cost constraint is satisfied already. We can repeat partitioning this set into two halves and sending messages in the directions of each half with equal probability to achieve the maximum rate $R =1$. Moreover, 
\begin{align}
    P[{Y_j^{(e)}} = 1 | \exists i \in[j-1]: \underline Y_i = (1,0) ] = 0,
    \label{eq:10}
    \\
    P[{Y_j^{(e)}} = 1 |  \underline Y_{j-1} = (1,1) ] = \frac{1}{2},
    \label{eq:10}
\end{align}
and to achieve $\underline Y_{j-1} = (1,1) $, the only possible sequences of $\underline Y^{j-2}$ are $(0^k 1^{j-2-k},0^k 1^{j-2-k})$ where $k\in[j-2]$ and $1^0 = \emptyset$. 
\end{itemize}
Therefore, 
to compute \eqref{eq:H1y}, 
we just need to focus on scenarios where $\underline y^{j-1} = (0^k 1^{j-1-k},0^k 1^{j-1-k})$ where $k\in[j-1]$ and $1^0 = \emptyset$. 
Based on Algorithm \ref{alg:JSAC-peak}, we obtain that
\begin{align*}
 & P_{ {Y^{(e)}}^{j-1}, {\underline S^{(e)}}}( 0^{j-1}, {\underline s^{(e)}}) = \frac{K - \sum_{k=1}^{j-1} c_k}{K^2},
 \notag 
 \\
   & P_{{Y_j^{(e)}} | {Y^{(e)}}^{j-1}, {\underline S^{(e)}}}(1 | 0^{j-1}, {\underline s^{(e)}}) \notag 
    \\& = \sum_{{y^{(l)}}^{j-1}} P_{{Y_j^{(e)}} | {Y^{(e)}}^{j-1}, {Y^{(l)}}^{j-1}, {\underline S^{(e)}}}(1 | 0^{j-1}, {y^{(l)}}^{j-1}, {\underline s^{(e)}}) \notag 
    \\& \qquad \times \frac{ P_{{Y^{(l)}}^{j-1}  {Y^{(e)}}^{j-1}, {\underline S^{(e)}}}({y^{(l)}}^{j-1}, 0^{j-1}, {\underline s^{(e)}}) }{  P_{ {Y^{(e)}}^{j-1}, {\underline S^{(e)}}}( 0^{j-1}, {\underline s^{(e)}})}\notag 
    \\& =  P_{{Y_j^{(e)}} | {Y^{(e)}}^{j-1}, {Y^{(l)}}^{j-1}, {\underline S^{(e)}}}(1 | 0^{j-1}, 0^{j-1}, {\underline s^{(e)}}) \notag 
    \\& \qquad \times \frac{ P_{{Y^{(l)}}^{j-1}  {Y^{(e)}}^{j-1}, {\underline S^{(e)}}}(0^{j-1}, 0^{j-1}, {\underline s^{(e)}}) }{  P_{ {Y^{(e)}}^{j-1}, {\underline S^{(e)}}}( 0^{j-1}, {\underline s^{(e)}})}\notag 
    \\& = \frac{c_j}{K - \sum_{k=1}^{j-1} c_k } \frac{K - \sum_{k=1}^{j-1} c_k}{K}  = \frac{c_j}{K},
    \\
    &P_{{Y^{(e)}}^{j-1}, {\underline S^{(e)}}}( 0^{j-2} 1, {\underline s^{(e)}})  \notag 
    \\& = P_{{Y^{(e)}}^{j-1}, {Y^{(l)}}^{j-1}, {\underline S^{(e)}}}( 0^{j-2} 1, 0^{j-2} 0,  {\underline s^{(e)}}) \notag
    \\& \qquad + P_{{Y^{(e)}}^{j-2}, {Y^{(l)}}^{j-1}, {\underline S^{(e)}}}( 0^{j-2} 1, 0^{j-2} 1 , {\underline s^{(e)}}) \notag 
    \\& = \frac{1}{K} \left(  \frac{c_{j-1} (K - \sum_{k=1}^{j-1} c_k) }{K^2} + \frac{c_{j-1}^2}{K^2}\right) \notag 
    \\& = \frac{c_{j-1}  (K - \sum_{k=1}^{j-2} c_k) }{K^3}, 
    \\
    &P_{{Y_j^{(e)}} | {Y^{(e)}}^{j-1}, {\underline S^{(e)}}}(1 | 0^{j-2}1, {\underline s^{(e)}}) \notag 
    \\& =  P_{{Y_j^{(e)}} | {Y^{(e)}}^{j-1}, {Y^{(l)}}^{j-1}, {\underline S^{(e)}}}(1 | 0^{j-2} 1, 0^{j-2} 1, {\underline s^{(e)}}) \notag 
    \\& \qquad \times \frac{ P_{{Y^{(l)}}^{j-1}  {Y^{(e)}}^{j-1}, {\underline S^{(e)}}}(0^{j-2} 1, 0^{j-2} 1, {\underline s^{(e)}}) }{  P_{ {Y^{(e)}}^{j-1}, {\underline S^{(e)}}}( 0^{j-2} 1, {\underline s^{(e)}})}\notag 
    \\& = \frac{1}{2 } \frac{c_{j-1}}{K - \sum_{k=1}^{j-2} c_k}, 
     \\ 
    &P_{{Y^{(e)}}^{j-1}, {\underline S^{(e)}}}( 0^{k} 1^{j-1-k}, {\underline s^{(e)}}) \ \ k <j-2  \notag 
    \\& = P_{{Y^{(e)}}^{j-1}, {Y^{(l)}}^{j-1}, {\underline S^{(e)}}}( 0^{k} 1^{j-1-k}, 0^{k} 1^{j-2-k} 0,  {\underline s^{(e)}}) \notag
    \\& \qquad + P_{{Y^{(e)}}^{j-2}, {Y^{(l)}}^{j-1}, {\underline S^{(e)}}}( 0^{k} 1^{j-1-k}, 0^{k} 1^{j-1-k} , {\underline s^{(e)}}) \notag 
    \\& = \frac{1}{K}  \frac{c_{k+1}^2}{K^2} (\frac{1}{2})^{2(j-k-2)-1}, 
    \\
    &P_{{Y_j^{(e)}} | {Y^{(e)}}^{j-1}, {\underline S^{(e)}}}(1 | 0^{k}1^{j-1-k}, {\underline s^{(e)}}) \notag 
    = \frac{1}{2 }, 
\end{align*}
where $i = \{1, \cdots, j-1\}.$ Now moving back to \eqref{eq:H1y}, we have
\begin{align}
 & \frac{1}{L} I( \underline X^L ; {Y^{(e)}}^L | {\underline S^{(e)}} ) \notag 
 \\&=  \frac{1}{L} \sum_{j=1}^L \sum_{  \underline s^{(e)}} \left( \sum_{k=1}^{j-1}  P_{ {Y^{(e)}}^{j-1}, {\underline S^{(e)}}}( 0^{k}1^{j-1-k}, {\underline s^{(e)}}) \right. \notag 
 \\&  \left. \qquad \times H( P_{{Y_j^{(e)}} | {Y^{(e)}}^{j-1}, {\underline S^{(e)}}}(1 | 0^{k}1^{j-1-k}, {\underline s^{(e)}}) )   \right)  \notag 
 \\& =  \frac{1}{L} \sum_{j=1}^L \left(  \frac{K - \sum_{k=1}^{j-1} c_k}{K} H(\frac{c_j}{K} ) \right. \notag 
 \\& \left. \qquad + \frac{c_{j-1}  (K - \sum_{k=1}^{j-2} c_k) }{K^2} H( \frac{1}{2 } \frac{c_{j-1}}{K - \sum_{k=1}^{j-2} c_k}  )  \right. \notag 
 \\& \left. \qquad + \sum_{k=1}^{j-3} \frac{1}{K}  \frac{c_{k+1}^2}{K^2} (\frac{1}{2})^{2(j-k-2)-1} 
 \right), \notag 
\end{align}
which is also the gap between the secrecy capacity inner and outer bounds. 
\end{IEEEproof}

The discrepancy between the derived outer and inner bounds provides insight into the remaining uncertainty regarding the true secrecy capacity. It is often observed that the gap tends to decrease with increasing $L$. This suggests that the JCAS scheme, which underpins the inner bound calculation, becomes increasingly efficient in approaching the theoretical outer limit as more in-block channel uses become available.

\section{Numerical Examples}
\label{sec:simulation}
\begin{figure}[h]
\centerline{ \includegraphics[width=1\linewidth]{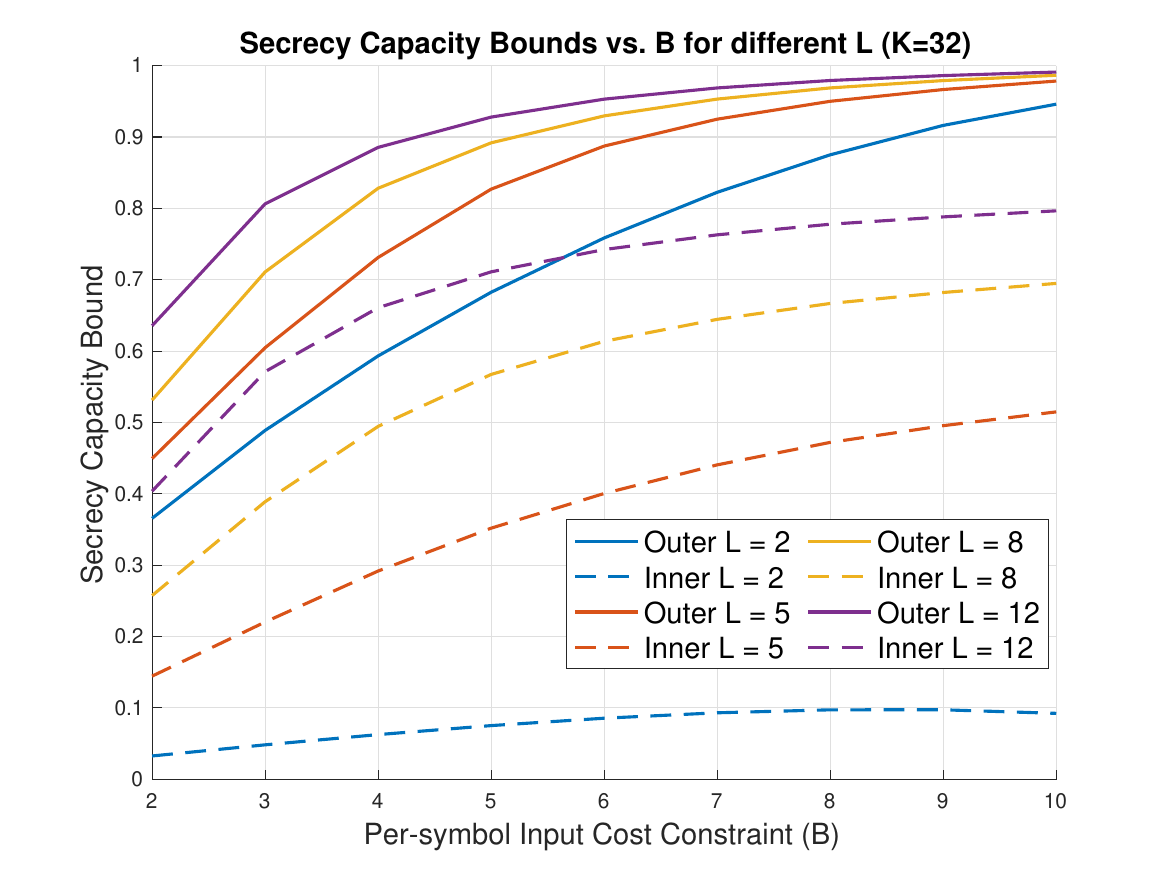} \hspace{.0cm}  }
\caption{Secrecy Capacity Inner and Outer Bounds.}
     \label{fig:example}
	\vspace{-.45cm}
\end{figure}

In this section, we present numerical results to illustrate the derived outer and inner bounds on the secrecy capacity of the BBP channel with feedback and iBM. We evaluate these bounds as a function of the per-symbol input cost constraint $B$ for various in-block memory lengths $L$. For the simulations, the number of possible quantized directions is set to $K=32$.
Specifically, Fig. \ref{fig:example} plots the secrecy capacity outer bound  in Theorem \ref{thm:upper} and the inner bound  in Theorem \ref{thm:lower} versus the input cost constraint $B$, for different values of $L \in \{2,5,8, 12 \}$. The outer bounds are depicted with solid lines, while the inner bounds are shown with dashed lines. Each color in the plot corresponds to a specific value of $L$.

The impact of the input cost constraint $B$ on the secrecy capacity bounds is evident across all considered values of $L$. Both the outer and inner bounds are observed to be non-decreasing functions of $B$. For smaller values of $B$, the secrecy capacity tends to increase significantly as $B$ rises. This behavior is expected, as a larger $B$ permits the exploration of a greater number of beam directions within each step of the beam selection process, defined by $c_j$ in \eqref{eq:b0s-peak}. This, in turn, enhances the legitimate user's channel estimation accuracy and subsequent data transmission capabilities. However, as $B$ continues to increase, for instance, when $B \geq K/2$ or $B$ exceeds the maximum possible $c_j$ for the initial steps, the rate of increase in secrecy capacity may diminish, eventually leading to saturation. This saturation occurs because the number of beams that can be effectively probed, particularly in the initial stages of a block, becomes limited by the total available directions $K$ (e.g., $K/2$) or the progressively decreasing set of unexplored directions, rather than by the constraint $B$ itself.

  Similarly, the in-block memory length $L$ plays a crucial role in determining the achievable secrecy capacity. For a fixed input cost constraint $B$, increasing $L$ generally results in higher secrecy capacity bounds for both the outer and inner limits. A larger $L$ provides additional opportunities to refine the beam alignment towards the legitimate user while simultaneously aiming to minimize information leakage to the eavesdropper over a greater number of transmissions. The improvement afforded by increasing $L$ is often more pronounced when moving from smaller values of $L$ (e.g., from $L=2$ to $L=5$) compared to transitions between already larger values (e.g., from $L=8$ to $L=12$), especially when $B$ is sufficiently large. 

\section{Conclusion}
\label{sec:conclusion}
This paper investigated the secrecy capacity of the binary beampointing (BBP) channel with block memory and feedback, a model relevant for secure beamformed transmissions. We established novel outer and inner bounds on this secrecy capacity. We derived a closed-form outer bound on the secrecy capacity by leveraging the capacity of the main channel. Furthermore, we established an achievable inner bound based on the Csiszár-Körner secrecy rate formula, adapted to the BBP channel with in-block memory. This inner bound is supported by a proposed joint communication and sensing (JCAS) transmission strategy, which dynamically adapts its beamforming based on the feedback received. Numerical evaluations demonstrated that both the input cost constraint $B$ and block length $L$ significantly influence the secrecy capacity, with larger values generally enhancing performance. These findings provide valuable insights into the fundamental trade-offs between communication reliability and information-theoretic secrecy in beamforming systems operating under adversarial conditions. The derived bounds and the proposed transmission scheme offer a theoretical basis for the design and optimization of secure JCAS systems.

\bibliographystyle{IEEEtran}
\bibliography{refs-isac}

\begin{thebibliography}{10}
\providecommand{\url}[1]{#1}
\csname url@samestyle\endcsname
\providecommand{\newblock}{\relax}
\providecommand{\bibinfo}[2]{#2}
\providecommand{\BIBentrySTDinterwordspacing}{\spaceskip=0pt\relax}
\providecommand{\BIBentryALTinterwordstretchfactor}{4}
\providecommand{\BIBentryALTinterwordspacing}{\spaceskip=\fontdimen2\font plus
\BIBentryALTinterwordstretchfactor\fontdimen3\font minus
  \fontdimen4\font\relax}
\providecommand{\BIBforeignlanguage}[2]{{%
\expandafter\ifx\csname l@#1\endcsname\relax
\typeout{** WARNING: IEEEtran.bst: No hyphenation pattern has been}%
\typeout{** loaded for the language `#1'. Using the pattern for}%
\typeout{** the default language instead.}%
\else
\language=\csname l@#1\endcsname
\fi
#2}}
\providecommand{\BIBdecl}{\relax}
\BIBdecl

\bibitem{6GTerahertz}
M.~Polese, J.~M. Jornet, T.~Melodia, and M.~Zorzi, ``Toward end-to-end,
  full-stack 6g terahertz networks,'' \emph{IEEE Communications Magazine},
  vol.~58, no.~11, pp. 48--54, 2020.

\bibitem{hong2021role}
W.~Hong, Z.~H. Jiang, C.~Yu, D.~Hou, H.~Wang, C.~Guo, Y.~Hu, L.~Kuai, Y.~Yu,
  Z.~Jiang \emph{et~al.}, ``The role of millimeter-wave technologies in 5g/6g
  wireless communications,'' \emph{IEEE Journal of Microwaves}, vol.~1, no.~1,
  pp. 101--122, 2021.

\bibitem{Michelusi_2018}
N.~Michelusi and M.~Hussain, ``Optimal beam-sweeping and communication in
  mobile millimeter-wave networks,'' in \emph{2018 IEEE International
  Conference on Communications (ICC)}, 2018, pp. 1--6.

\bibitem{ISAC-6G}
F.~Liu, Y.~Cui, C.~Masouros, J.~Xu, T.~X. Han, Y.~C. Eldar, and S.~Buzzi,
  ``Integrated sensing and communications: Towards dual-functional wireless
  networks for 6g and beyond,'' \emph{IEEE Journal on Selected Areas in
  Communications}, pp. 1--1, 2022.

\bibitem{EnableJCAS}
J.~A. Zhang, M.~L. Rahman, K.~Wu, X.~Huang, Y.~J. Guo, S.~Chen, and J.~Yuan,
  ``Enabling joint communication and radar sensing in mobile networks—a
  survey,'' \emph{IEEE Communications Surveys \& Tutorials}, vol.~24, no.~1,
  pp. 306--345, 2022.

\bibitem{ISAC-IT}
M.~Ahmadipour, M.~Kobayashi, M.~Wigger, and G.~Caire, ``An
  information-theoretic approach to joint sensing and communication,''
  \emph{IEEE Transactions on Information Theory}, pp. 1--1, 2022.

\bibitem{Li2022Asilomar}
S.~Li and G.~Caire, ``On the capacity of “beam-pointing” channels with
  block memory and feedback: The binary case,'' in \emph{2022 56th Asilomar
  Conference on Signals, Systems, and Computers}, 2022, pp. 1262--1268.

\bibitem{Li2023ISIT}
------, ``On the capacity and state estimation error of binary "beam-pointing"
  channels with block memory and feedback,'' in \emph{2023 IEEE International
  Symposium on Information Theory (ISIT)}, 2023, pp. 2571--2576.

\bibitem{BBP:TIT2023}
------, ``On the capacity and state estimation error of “beam-pointing”
  channels: The binary case,'' \emph{IEEE Trans. Inf. Theory}, vol.~69, no.~9,
  p. 5752–5770, Sep. 2023.

\bibitem{Huang2012TIT}
K.~Huang and V.~K.~N. Lau, ``Stability and delay of zero-forcing sdma with
  limited feedback,'' \emph{IEEE Transactions on Information Theory}, vol.~58,
  no.~10, pp. 6499--6514, 2012.

\bibitem{Li2019ITW}
S.~Li, H.~Seferoglu, D.~Tuninetti, and N.~Devroye, ``On the stability region of
  the layered packet erasure broadcast channel with output feedback,'' in
  \emph{2019 IEEE Information Theory Workshop (ITW)}, 2019, pp. 1--5.

\bibitem{Li2019ICC}
S.~Li, D.~Tuninetti, and N.~Devroye, ``On the capacity region of the layered
  packet erasure broadcast channel with feedback,'' in \emph{ICC 2019 - 2019
  IEEE International Conference on Communications (ICC)}, 2019, pp. 1--6.

\bibitem{Tatiknonda2009}
S.~Tatikonda and S.~Mitter, ``The capacity of channels with feedback,''
  \emph{IEEE Transactions on Information Theory}, vol.~55, no.~1, pp. 323--349,
  2009.

\bibitem{Li2021ISIT}
S.~Li, D.~Tuninetti, and N.~Devroye, ``A control-theoretic linear coding scheme
  for the fading gaussian broadcast channel with feedback,'' in \emph{2021 IEEE
  International Symposium on Information Theory (ISIT)}, 2021, pp. 1--6.

\bibitem{caire2007multiuser}
G.~Caire, N.~Jindal, M.~Kobayashi, and N.~Ravindran, ``Multiuser mimo downlink
  made practical: Achievable rates with simple channel state estimation and
  feedback schemes,'' \emph{Arxiv preprint cs. IT}, vol. 710, 2007.

\bibitem{Li2023ITW}
S.~Li and G.~Caire, ``On the state estimation error of "beam-pointing"
  channels: The binary case,'' in \emph{2023 IEEE Information Theory Workshop
  (ITW)}, 2023, pp. 430--434.

\bibitem{wiretap}
A.~D. Wyner, ``The wire-tap channel,'' \emph{The Bell System Technical
  Journal}, vol.~54, no.~8, pp. 1355--1387, 1975.

\bibitem{Csiszar-Korner}
I.~Csiszar and J.~Korner, ``Broadcast channels with confidential messages,''
  \emph{IEEE Transactions on Information Theory}, vol.~24, no.~3, pp. 339--348,
  1978.

\bibitem{secrecy-fading-2008}
P.~K. Gopala, L.~Lai, and H.~El~Gamal, ``On the secrecy capacity of fading
  channels,'' \emph{IEEE Transactions on Information Theory}, vol.~54, no.~10,
  pp. 4687--4698, 2008.

\bibitem{wang2007secrecy}
P.~Wang, G.~Yu, and Z.~Zhang, ``On the secrecy capacity of fading wireless
  channel with multiple eavesdroppers,'' in \emph{2007 IEEE International
  Symposium on Information Theory}.\hskip 1em plus 0.5em minus 0.4em\relax
  IEEE, 2007, pp. 1301--1305.

\bibitem{shafiee2007achievable}
S.~Shafiee and S.~Ulukus, ``Achievable rates in gaussian miso channels with
  secrecy constraints,'' in \emph{2007 IEEE International Symposium on
  Information Theory}.\hskip 1em plus 0.5em minus 0.4em\relax IEEE, 2007, pp.
  2466--2470.

\bibitem{secureMISOME2010}
A.~Khisti and G.~W. Wornell, ``Secure transmission with multiple antennas i:
  The misome wiretap channel,'' \emph{IEEE Transactions on Information Theory},
  vol.~56, no.~7, pp. 3088--3104, 2010.

\bibitem{yeh2021eavesdropping}
C.-Y. Yeh and E.~W. Knightly, ``Eavesdropping in massive mimo: New
  vulnerabilities and countermeasures,'' \emph{IEEE Transactions on Wireless
  Communications}, vol.~20, no.~10, pp. 6536--6550, 2021.

\bibitem{lai2008wiretap}
L.~Lai, H.~El~Gamal, and H.~V. Poor, ``The wiretap channel with feedback:
  Encryption over the channel,'' \emph{IEEE Transactions on Information
  Theory}, vol.~54, no.~11, pp. 5059--5067, 2008.

\bibitem{bashar2011secrecy}
S.~Bashar, Z.~Ding, and G.~Y. Li, ``On secrecy of codebook-based transmission
  beamforming under receiver limited feedback,'' \emph{IEEE transactions on
  wireless communications}, vol.~10, no.~4, pp. 1212--1223, 2011.

\bibitem{permuter2008capacity}
H.~Permuter, P.~Cuff, B.~Van~Roy, and T.~Weissman, ``Capacity of the trapdoor
  channel with feedback,'' \emph{IEEE Transactions on Information Theory},
  vol.~54, no.~7, pp. 3150--3165, 2008.

\bibitem{gunlu2023secure}
O.~G{\"u}nl{\"u}, M.~R. Bloch, R.~F. Schaefer, and A.~Yener, ``Secure
  integrated sensing and communication,'' \emph{IEEE Journal on Selected Areas
  in Information Theory}, vol.~4, pp. 40--53, 2023.

\end{thebibliography}

\end{document}